\documentclass[]{rmf-d}
\usepackage{nopageno,rmfbib,multicol,times,epsf,amsmath,amssymb} % En inglés cambiar spanish por english
\usepackage[latin1]{inputenc} % comentar para Inglés
\usepackage[]{caption2}
\usepackage{graphics}

\def\rmfcornisa{INVESTIGACI\'{O}N \hfill\rmf\ {\bf 48} (6) ???--??? % Suponiendo Vol 48 num 6 y area ENSEÑANZA
\hfill Febrero 2007}

\def\rmfcintilla{{\it Rev.\ Mex.\ F\'{\i}s.\/} {\bf ??} (?) (2007) ???--???}
\clearpage \rmfcaptionstyle 
\begin{document}
\title{Formaci\'{o}n del ion negativo de carb\'{o}n por asociaci\'{o}n
radiativa
\vspace{-6pt}}

\author{J. Campos, A. Lipovka, J. Saucedo}
\address{Centro de Investigaci\'{o}n en F\'{i}sica, Universidad de Sonora\\
Rosales y Blvd. Transversal, Col. Centro, Edif. 3-I 83000 Hermosillo,
Sonora, MEXICO\\
Apartado Postal 5-088 Tel.: (52-662) 259-21-56. Fax: (52-662) 212-66-49.
E-mail: jcampos@cajeme.cifus.uson.mx}
\author{ V. Zalkind}
\address{Physical Tecnical Institute de Ioffe\\
26 polytekhnicheskaya, St Petersburg 194021, Russian Federation\\
Fax: (812) 297-10-17, Phone: (812) 297-10-17}

\maketitle \recibido{ de  de }{ de  de \vspace{-12pt}} % Fecha de recibido y aceptado (ej. 3 de abril de 2002)

\begin{resumen}
En el presente trabajo es considerado el problema de la asociaci\'{o}n
radiativa de \'{a}tomos de carb\'{o}n C con electrones $e^{-}$ en un
intervalo de temperaturas cin\'{e}ticas $T_{c}$ de $100 < T_{c} < 3000$
K. El c\'{a}lculo de la raz\'{o}n de dicha asociaci\'{o}n se ha
hecho, empleando el principio del balance detallado. Se muestra que la raz%
\'{o}n se comporta de manera correcta conforme la temperatura se incrementa.
Sucede algo por el estilo, respecto a la formaci\'{o}n del ion negativo del
hidr\'{o}geno $H^{-}$, donde la raz\'{o}n tambi\'{e}n crece con la
temperatura.
% Incluir aqui el resumen
\end{resumen}

\descript{ Procesos moleculares y qu\'{i}micos e interacciones;
plasma astrof\'{i}sico; cosmolog\'{i}a \vspace{-4pt}}

\begin{abstract}
In the present paper the problem of the radiative association of atoms of
carbon C with electrons $e^{-}$ at an interval of kinetic temperatures
$T_{c}$ of $100 < T_{c} < 3000$ K is considered. The calculation
of the rate of these associations has been made by using the principle of
detailed balance. It is shown that the rate has correct behavior (it
increase with the temperature) it is behavior is look-like that for the
$H^{-}$ formation rate coefficient, which also increase with the temperature.
\end{abstract}

\keys{Molecular and chemical processes and interactions;
astrophysical plasma; cosmology \vspace{-4pt}} \pacs{M 95.30.Ft; 95.30.Qd; 98.80.-k \vspace{-4pt}}

\begin{multicols}{2}
\section{Introducci\'{o}n}
En los \'{u}ltimos a\~{n}os ha aparecido en la literatura, mucho inter\'{e}s
sobre la formaci\'{o}n de mol\'{e}culas de $CH$ en la cosmolog\'{i}a. Este
inter\'{e}s, es debido a que \'{a}tomos de carb\'{o}n, nitr\'{o}geno y
ox\'{i}geno estan formando un pico en la distribuci\'{o}n de n\'{u}cleos que
se forman en n\'{u}cleos pr\'{i}migenios. Estas considerables abundancias de
mol\'{e}culas basadas en estos \'{a}tomos pesados, pueden ser los \'{u}nicos
instrumentos para medir condiciones f\'{i}sicas que aparecen en el universo
temprano y que nos permiten elegir entre diferentes modelos de
nucleos\'{i}ntesis primordial no est\'{a}ndar (los cuales se discuten mucho
ahora). Estas mismas abundancias, tambi\'{e}n se les puede usar para
establecer l\'{i}mites superiores a las abundancias predichas por el modelo
est\'{a}ndar de nucleos\'{i}ntesis primordial.

El ion negativo de carb\'{o}n $C^{-}$, el cual tiene una energ\'{i}a de
amarre de $D_{0}=1.25$ $eV$, es m\'{a}s estable que el ion negativo de
hidr\'{o}geno $H^{-}$ con $D_{0}=0.75$ $eV$, y esta jugando un papel
importante en qu\'{i}mica del carb\'{o}n en combustibles y tambi\'{e}n en
qu\'{i}mica del gas en astrof\'{i}sica (cascaras de estrellas, de SuperNovas
(SN), cin\'{e}tica molecular de nubes moleculares en la galaxia, asi como
extragal\'{a}cticos).

La mol\'{e}cula $CH$ juega un papel principal, como es citado en nuestro
art\'{i}culo anterior [1], debido a que por un lado se forma m\'{a}s
r\'{a}pido durante la \'{e}poca pregal\'{a}ctica y por otro lado el
carb\'{o}n $C$ es una especie muy sensible a los modelos de Big Bang y a
inhomogeneidades primigenias. Como fue mencionado en [1], los canales
principales de formaci\'{o}n de $CH$ primordial son: $C+H_{2}\rightarrow CH+H
$, pero, como en el caso del hidr\'{o}geno molecular $H_{2}$, el cual se
forma por medio de la cadena: $H+e^{-}\rightarrow H^{-}+\gamma $ y $%
H^{-}+H\rightarrow H_{2}+e^{-}$ formando iones negativos de $H$, en la
formaci\'{o}n molecular de $CH$ deber\'{i}a jugarse un canal con ayuda de un
ion negativo de carb\'{o}n $C^{-}$. Esta especie se forma por medio de
la reacci\'{o}n $C+e^{-}\rightarrow C^{-}+\gamma $. La raz\'{o}n de esta
reacci\'{o}n fue calculada en [2]. Pero, en tal art\'{i}culo los autores
usaron una aproximaci\'{o}n de secciones eficaces para energ\'{i}as
bajas, la cual cuenta s\'{o}lo con un t\'{e}rmino que corresponde a las
velocidades peque\~{n}as de las especies. Tal hecho, los llevo a una
raz\'{o}n la cual tiene un comportamiento incorrecto en el caso de las
temperaturas de inter\'{e}s (desde $100K$ a $1000K$), las cuales se asocian al
caso de formaci\'{o}n molecular a trav\'{e}s de la \'{e}poca oscura. Por eso es
muy importante calcular dicha taza.

En consecuencia del principio del balance detallado, la secci\'{o}n eficaz
del proceso puede ser recalculado con la ayuda de la secci\'{o}n eficaz
correspondiente a un proceso inverso [3,4]. Tal proceso inverso es el
desprendimiento radiativo $C^{-}+\gamma \rightarrow C+e^{-}$.

En este art\'{i}culo estamos calculando la raz\'{o}n de asociaci\'{o}n
radiativa de $C$ con $e^{-}$, formando un ion negativo de carb\'{o}n $%
C^{-}$ usando el principio del balance detallado.

\section{Ecuaci\'{o}nes para el C\'{a}lculo de la Raz\'{o}n}

El proceso de desprendimiento fue considerado en detalle en [5]. En este
art\'{i}culo fue calculada la secci\'{o}n eficaz del proceso de
desprendimiento de $C^{-}$, y el c\'{a}lculo coincide muy bien con el
experimento.

Consideramos un balance detallado entre reacciones de ambos lados, directa e
inversa. El n\'{u}mero de cantidad de reacciones de la asociaci\'{o}n
radiativa en un $cm^{3}/s$ en un intervalo de velocidades desde $v$ hasta $%
v+dv$ esta dado por

\begin{equation}
Z_{a}=N_{C}N_{e}f\left( v\right) vdv\sigma _{A},\text{ }
\end{equation}

\noindent donde la funci\'{o}n de distribuci\'{o}n de Maxwell integrada por los
\'{a}ngulos $\theta $ y $\phi $ es

\begin{equation}
f\left( v\right) =4\pi v^{2}\left( \frac{m}{2\pi kT}\right) ^{3/2}\exp
\left( \frac{-mv^{2}}{2kT}\right) ,\text{ }
\end{equation}

\noindent donde $v$ es la velocidad relativa entre dos especies, $N_{C}$ y $N_{e}$ son
abundancias de $C$ y $e^{-}$. Por otro lado, el n\'{u}mero de reacciones
reversas (desprendimiento radiativo) en un $cm^{3}/s$ en el intervalo
de frecuencias desde $\nu $ hasta $\nu +d\nu $ esta dado por

\begin{equation}
Z_{d}=N_{nC^{-}}\frac{U_{\nu }}{h\nu }d\nu c\sigma _{d}\left[ 1-\exp \left(
\frac{-h\nu }{kT}\right) \right] ,
\end{equation}

\noindent donde $N_{nC^{-}}$ es la abundancia de $C^{-}$ y $U_{\nu }$ es la densidad
de energ\'{i}a de cuerpo negro, la cual es dada por

\begin{equation}
U_{\nu }=\frac{8\pi h\nu ^{3}}{c^{3}}\frac{1}{\left[ \exp \left( \frac{h\nu
}{kT}\right) -1\right] },
\end{equation}

\noindent donde $c$ es la velocidad de la luz.

La densidad de $C^{-}$ esta dada por la relaci\'{o}n

\begin{equation}
N_{nC^{-}}=\frac{g_{m}}{Z_{C^{-}}}N_{C^{-}}\exp \left( \frac{-E_{m}+D_{0}}{kT%
}\right) ,
\end{equation}

\noindent donde $D_{0}$ es la energ\'{i}a de desprendimiento, $g_{m}$ es el peso
estad\'{i}stico, $Z_{C^{-}}$ es la suma estad\'{i}stica y $N_{C^{-}}$ es la
abundancia total de $C^{-}$.

Igualando las ecuaciones (1) y (3) y usando la ecuaci\'{o}n (5), obtenemos
la relaci\'{o}n para secciones eficaces para el des\-pren\-di\-mien\-to radiativo

\begin{eqnarray}
\sigma _{a}=\sigma _{d}\frac{g_{m}}{Z_{C^{-}}}\frac{U_{\nu }}{f\left(
v\right) }\left( \frac{d\nu }{dv}\right) \frac{c}{h\nu v}\frac{N_{C^{-}}}{%
N_{C}N_{e}} \times \mbox{} \nonumber \\
\left[ 1-\exp \left( \frac{-h\nu }{kT}\right) \right] \exp \left(
\frac{-E_{m}+D_{0}}{kT}\right) .
\end{eqnarray}
%% \mbox{} \nonumber \\

En consecuencia con la ecuaci\'{o}n de Saha, las abundancias son de la forma

\begin{equation}
\frac{N_{C^{-}}}{N_{C}N_{e}}=\left( \frac{h^{2}}{2\pi mkT}\right) ^{3/2}%
\frac{Z_{C^{-}}}{Z_{C}Z_{e}}\exp \left( \frac{D_{0}}{kT}\right) .
\end{equation}

Introduciendo (7) en (6) y tomando en cuenta la ley de conservaci\'{o}n de
la energ\'{i}a

\begin{equation}
h\nu =\frac{mv^{2}}{2}-E_{n}.
\end{equation}

Entonces, tenemos la relaci\'{o}n de secciones eficaces

\begin{equation}
\sigma _{a}=\frac{2g_{n}}{Z_{C}Z_{e}}\left( \frac{h\nu }{mcv}\right)
^{2}\sigma _{d},
\end{equation}

\noindent donde $Z_{C}$ es la suma estad\'{i}stica.

Finalmente la raz\'{o}n de la asociaci\'{o}n radiativa $C+e\rightarrow
C^{-}+\gamma $, esta dada por la integral

\begin{equation}
R_{a}\left( T_{c}\right) =\int_{0}^{\infty }\sigma _{a}f\left( v\right) vdv.
\end{equation}

\noindent donde $f\left( v\right) $ es la funci\'{o}n de distribuci\'{o}n dada por la
ec. (2) y $v$ es la velocidad del electr\'{o}n.

\section{Resultados de C\'{a}lculo}

Las relaciones (9) y (10) con los datos para la secci\'{o}n eficaz del
proceso inverso ($C^{-}+\gamma \rightarrow C+e^{-}$) [5] nos permite
resolver la tarea y calcular la secci\'{o}n eficaz y la raz\'{o}n para la
asociaci\'{o}n radiativa $C+e^{-}\rightarrow C^{-}+\gamma $. A diferencia
del c\'{a}lculo realizado en [2], nosotros realizamos el c\'{a}lculo tomando
en cuenta el principio del balance detallado y as\'{i} obtener tal raz\'{o}n con
el comportamiento m\'{a}s correcto en la regi\'{o}n de temperaturas de
inter\'{e}s, comparandolo con resultados de [2].

%\begin{minipage}[t]{.25\linewidth}
%\centering
%\includegraphics[width=3.2in]{figcocc.eps}
%\end{minipage}%
\begin{center}
\begin{minipage}{9cm}
\includegraphics{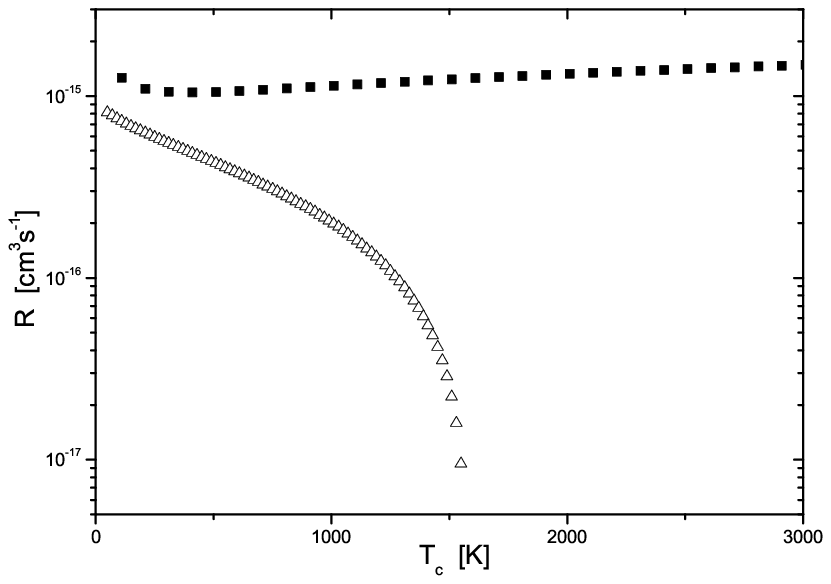}
\end{minipage}
\small{Figura 1. Resultado de la raz\'{o}n $R_{a}\left(T_{c}\right)$ para la formaci\'{o}n de iones negativos de carb\'{o}n $C^{-}$ usando el principio del balance detallado. La curva de cuadritos negros representa nuestra predicci\'{o}n te\'{o}rica. Los triangulos blancos describen el resultado obtenido en [2].}
\end{center}

%\begin{center}
%\includegraphics[width=.5in]{figcocc.eps}
%\end{center}
%\includegraphics{scale=0.5}{figcocc.eps}

En la figura 1 mostramos el resultado de nuestro c\'{a}lculo, junto con el
obtenido en [2]. Dentro de la figura, los cuadritos negros representan a
nuestro resultado, el cual fue obtenido usando el principio del balance
detallado. Los triangulos blancos representan el resultado ofrecido en [2].
La comparaci\'{o}n de ambos resultados muestra un buen acuerdo de los
c\'{a}lculos en regiones de baja temperatura cin\'{e}tica $T_{c}$. Sin
embargo, conforme dicha temperatura se incrementa, empiezan a ser visibles
las discrepancias de ambas predicciones te\'{o}ricas para la raz\'{o}n. Se
ve claramente que mientras la temperatura cin\'{e}tica crece, la raz\'{o}n
ofrecida en [2] se cae fuertemente. Tal comportamiento (como han mencionado
los autores) aparece debido a una simplificaci\'{o}n en sus c\'{a}lculos, es
decir, cuando ellos cancelaron la parte de secci\'{o}n eficaz que
corresponde a las energ\'{i}as altas. Dicha simplificaci\'{o}n realmente
no afect\'{o} la raz\'{o}n en el r\'{e}gimen de tem\-pe\-ra\-tu\-ras peque\~{n}as, pero
en el caso de temperaturas altas y medias, la discrepancia se aumenta mucho.
Tales discrepancias ilustran la descripci\'{o}n correcta predicha en el
l\'{i}mite de altas temperaturas por nuestro c\'{a}lculo, las cuales podemos
comparar con la bien conocida raz\'{o}n para formaci\'{o}n de $H^{-}$ ($%
H+e^{-}\rightarrow H^{-}+\gamma $), la cual es muy parecida a nuestro
resultado.

\section{Conclusi\'{o}n}

En este trabajo se deriv\'{o} una expresi\'{o}n anal\'{i}tica que relaciona a
las secciones eficaces de asociaci\'{o}n y de des\-pren\-di\-mien\-to (ver ec. 9) a
partir del principio del balance detallado. Posteriormente, calculamos la
raz\'{o}n de asociaci\'{o}n radiativa $R_{a}$ de \'{a}tomos de carb\'{o}n $C$
con electrones $e^{-}$, a trav\'{e}s de la reacci\'{o}n $C+e^{-}\rightarrow
C^{-}+\gamma $ y usando tal principio. Nuestro procedimiento de
obtenci\'{o}n de la raz\'{o}n de asociaci\'{o}n result\'{o} ser bastante
correcto, en un amplio intervalo de temperaturas, tanto bajas como altas
(desde $100^{\circ }K$ hasta $3000^{\circ }K$), si comparamos con la misma
raz\'{o}n de formaci\'{o}n de $H^{-}$.

%
%para indicar alguna figura, incluya el comentario "%figura_(numero de figura)",
%favor de incluir en la información del artículo los archivos correspondientes a las figuras en .bmp o en .jpg

\end{multicols}
\medline
\begin{multicols}{2}

% Incluir Aqui la Bibliografía
%\end{thebibliography}
\end{multicols}

\begin{thebibliography}{99}
\bibitem{Li} A. Lipovka, J. Saucedo, J. Campos, RMF 48 (2002) 325-334.

\bibitem{Ja} R.K. Janev, H.V. Regemorte, Astron. \& Astrophys., 37, 1-6 (1974).

\bibitem{Ma} J.K. Martin, Physical Review, 97 (1955) 1446.

\bibitem{Co} F. Coester, Letter to the Editor, (1951).

\bibitem{Mo} Yu.V. Moskvin, Opt. Spektrosk., 17, 270 (1964).
\end{thebibliography}
\end{document}